# Millimeter Wave Multi-Beam-Switching Antenna

Vedaprabhu Basavarajappa[1], Beatriz Bedia Exposito[1], Lorena Cabria[1] and Jose Basterrechea[2]

[1]Dept. of Antennas, TTI Norte
Parque Científico y Tecnológico de Cantabria,
C/ Albert Einstein nº 14, 39011
Santander, Spain

[2] Departamento de Ingeniería de Comunicaciones, Universidad de Cantabria
Edificio Ingeniería de Telecomunicación Profesor José Luis García García
Plaza de la Ciencia, Avda de Los Castros, 39005
Santander, Spain
Email: {veda, bbedia, lcabria} @ttinorte.es, jose.basterrechea@unican.es

*Abstract*—The paper introduces a high gain end-fire printed antenna operating at 14 GHz with a bandwidth of 2 GHz. The 3x3 antenna sub-array formed of this element is used to present its beam steering and multi-beam-switching capabilities for millimeter wave Massive MIMO antenna arrays of 5G networks. The co-existence of these two techniques in a Massive MIMO sub-array, aided by the use of a simple two vector excitation matrix technique, is shown in this paper. The high gain and small form factor of the antenna is useful in applications, like 5G arrays, where space is premium and the inherent layered structure also aids to the better cooling of the antenna front end.

## I. INTRODUCTION

Recent trends in telecommunications have shown an accelerated growth in the cellular data rate. According to the CISCO Visual Networking Index (VNI) of 2017, global mobile data traffic grew by 63 percent in 2016. Although 4G connections represented a quarter of the mobile connections, they accounted for 69 percent of the mobile data traffic. Therefore, in the recent years trends in 4G hint at an increasing demand for high data rates and QoS, that serves as the impetus for the 5G domain. In current cellular network architecture, the mobile data traffic is exchanged for most part over the air (OTA) between the cellular base station (eNodeB) and the mobile handset (UE). Antennas play a pivotal role in this link and this has recently triggered a lot of research in this area. One of the fundamental technologies that can help to achieve these data rates is the Massive MIMO technology. It involves the use of several (hundreds) of antennas to achieve high beamforming gains, and therefore, better peak throughput. The antennas that go into the core of a Massive MIMO setup for 5G networks have to meet the criterion of high gain, wide bandwidth, and preferably, small volume, which, by theory are not mutually exclusive factors themselves [1][2]. The requirement of large spectrum of bandwidth can be addressed using low volume antennas by operating them in the millimeter wave band, which offers a wide bandwidth.

Two Massive MIMO setups that have been tested successfully are the Argos [3] and the Lund LuMaMi [4]. The Argos has a large number of antennas that are deployed at the base station, which implies that the baseband processing block, the channel estimation, transmission synchronization and the channel calibration need to be scaled up too which can be challenging. It demonstrates a platform with 64 antennas operating at 2.4 GHz. The LuMaMi [4] antenna system consists of a T-shaped array made of λ/2 dual polarized shorted patch elements operating at 3.7 GHz. The feed placement and element size vary from the center of the array to the edges, to compensate for the array effects, that influence individual elements differently. Both of these antenna system setups operate in the below 6 GHz microwave region. Despite the benefits offered by these systems, it can be seen from their setups that, owing to the large operating wavelengths of the individual antenna, the antenna array front end of these systems can get large and bulky in order to attain a high gain. Using the millimeter wave band decreases, the size of the antenna and consequently a large number of antennas can be packed in a small area in an array to obtain high gain beams [5]. Antennas are designed for operation in the mmWave band, providing many additional functions. A tilted combine beam antenna operating at 28 GHz with a gain of 7.4 dBi is presented in [6], which provides a tilt in the elevation plane without the use of antenna arrays. An inkjet printed multilayer Yagi Uda antenna is presented in [7], which has the major advantage of post process deposition of antennas onto any rigid or flexible active/passive circuit topology. It can be seen this brings in an extra degree of freedom, namely blending, into the pre-existing design. The antenna has a gain of 8 dBi and operates in the 21-27 GHz band. Since millimeter wave and Massive MIMO seek high gain antennas that can be beam steered efficiently with a narrow beamwidth [8], it would be coherent to design these high gain antenna arrays using high gain individual array element antennas in the millimeter wave band that, constructively, add up to provide the final high array gain. Owing to their low profile, low cost and ease of integration, end-fire antennas are a good choice in this regard.

The end-fire antennas that are most commonly used in literature are the antipodal Vivaldi antenna, helical antenna and the Yagi-Uda antenna. Antipodal Vivaldi antenna has been used in applications such as UWB, phased array, radars and microwave imaging. Helical antennas have been recently used in applications that require circular polarization and, also, in mobile satellite communications. The Yagi-Uda antenna, which consists of a driven element supported by reflector and director elements [9], has seen wide use in television broadcast and has recently been used as a printed antenna as well. The Yagi- Uda antenna, in its original form, was a wire antenna and was proposed as the projector of the sharpest beam of electric waves by Yagi and Uda. The advantages of using the printed version are the mechanical stability, that is offered to the antenna due to the substrate and, also, the compactness that can be achieved by selecting a higher permittivity substrate. The compactness of resonant type antennas and the broad bandwidth characteristics were both simultaneously achieved in a single Yagi antenna, as described in [10]. The antenna, in this case, is fed by a microstrip to CPS transition balun, which, in a way, adds to the profile length. With the compactness and broad bandwidth addressed by antennas in [9] and [10], papers [11] and [12] focus on improving the antenna gain using techniques of a microstrip fed inset patch as the driver with parasitic patches added in front of the driving patch as directors to enhance the gain. Since Yagi-Uda antennas or the endfire antennas in general are protrude-out-of-wall configurations, it adds immensely if this dimension of profile length could be minimized while retaining the high gain characteristics. The antenna that is presented here offers the characteristics of [9]-[12] and further improves upon them, by having a different feeding structure to minimize the profile length and a different dipole driver, so as to better launch the surface waves. The dipole feeding point is based upon [13], where the antenna operates in the 2-4 GHz band with a gain around 5 dBi over the band.

In this work, an antenna is designed to operate in the 13.5 GHz to 15.5 GHz and there is an improvement in gain of additional 3 dBi by incorporating a specific novel arrangement of arc shaped parasitics in front of the driver dipole. This helps to attain the high gain with the two arms of the driver dipole printed on opposite sides of the substrate.

The antenna presented in this paper is used in a 3x3 subarray. A 4x4 subarray will finally be scaled to 8x8 in the final prototype. The beam steering capabilities of the antenna are shown by exciting a row of the subarray progressively in phase to steer the beam. In addition, simultaneous multiple beams are cast in several beamstates, giving the antenna the potential for use in multi-user scenarios in Massive MIMO.

The paper is organized as follows: Section I introduces, Section II gives a system level overview of the planned setup with the antenna system, Section III details the antenna design and its characteristics, Section IV provides the beam steering characteristics of the antenna array and presents the novel two vector beam switching method to generate simultaneous high gain multiple beams and Section V concludes the discussion.

## II. SYSTEM LEVEL DEFINITION

Along the way to realize 5G we propose that antennas need to be designed to cater to the needs of the three prominent technologies that are driving 5G, namely the Massive MIMO, the mmWave beamforming and the Single RF MIMO. While we propose that a single antenna system can be designed to meet the specifications of all three technologies giving the advantage of cross functional adaptability, we do not exclude the possibility of designing tailored antennas individually for each of these. Massive MIMO antennas are large in number (generally > 50), operate mostly in the sub 6 GHz band and are large in size, as in the test beds discussed in the introduction. Millimeter wave beamforming, on the other hand, also involves large number of antennas but, owing to the short wavelength, occupies a low volume. Millimeter wave frequencies in this paper will refer to frequencies in the range 10 GHz to 300 GHz [14]. Single RF MIMO antenna has been also investigated in the sub 6 GHz range extensively. The papers [15], [16] provide good examples of the realization. A Single RF MIMO antenna design at 15 GHz would easily follow from this. Recently, there has been increased interest in the 15 GHz frequency spectrum. Ericsson and NTT Docomo have carried out trials of 5G radio access in the 15 GHz band over a system bandwidth of 400 MHz [17] [18]. Nokia has considered the 15 GHz band with 64 antennas as an example scenario [19].

Therefore, building on this background, it could be said that Massive MIMO, mmWave and Single RF MIMO could all be realized over the 14 GHz spectrum (bandwidth 2 GHz), with antennas designed to this frequency. The final system level architecture in the model would include 64 antennas in the Massive MIMO regime, 3x3 antennas for the mmWave regime and a single RF 3x3 MIMO antenna. The building antenna block of the three technologies would be the common end-fire antenna operating at 14 GHz. The multi beams required by the Massive MIMO setup would be realized by a two-vector antenna excitation matrix. The additional beam steering required by the mmWave will be realized with progressive phase shift along rows of the array. The single RF MIMO requirement would be met by casting multiple beamstates through the same antenna spread over time (TDD), thereby minimizing the number of RF chains. This concept of interdependency is captured in the Venn diagram shown in Figure 1. The printed end-fire element which is at the region of convergence of the Venn diagram is of particular interest, owing to its two important properties: the presence of an endfire mechanism in the antenna, giving it a high gain as an individual element itself, and the possibility to carry out single RF parasitic antenna array setup integrated into the printed antenna. This element could further be used as a sub-array in a Massive MIMO array.

The aim of this paper is two-fold: the operation of a printed antenna that projects a high gain end fire beam and the demonstration of its use in beamforming in the millimeter wave frequencies by computer aided modeling and simulation. Further, an attempt will be made to highlight the synergy among Massive MIMO, Single RF and millimeter wave beamforming, by presenting a single simulation model that demonstrates the property of each of these technologies as a pre-requisite for further work in this area.

The simulation model of the antenna and its results with its evaluation for integrated Massive MIMO is described. The millimeter wave beam steering by varied phases is also shown. The end fire antenna element is used in a 3x3 antenna array. The elements in the antenna array are excited by amplitude and phase pattern, as described by the Table 1, and the depicted matrices. Patterns with multi-beams of high directivity in the boresight, and simultaneous beams projected at various angles are obtained through simulations.

This highlights the millimeter wave beamforming performed in a Massive MIMO sub array. Single RF MIMO will be targeted using two techniques: using the parasitic elements in the end fire antenna itself, thereby highlighting their interdependency using a single model and by casting beamstates spread in time (TDD). All simulations were carried out in Ansoft HFSS. The antenna model and the array antenna model were simulated using the full-wave solver in HFSS.

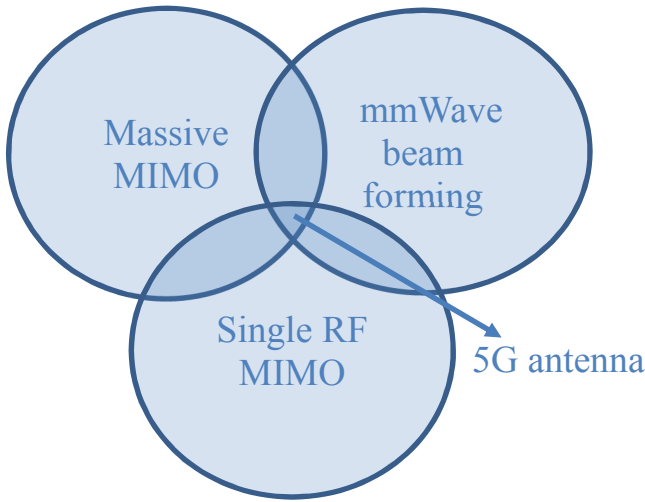

Figure 1. Venn diagram of interdependency

### III. ANTENNA DESIGN

The antenna design shown in Figure 2 consists of a printed end fire configuration with additional director elements to enhance the directivity along the end fire direction. The antenna substrate measures 12 mm x 11 mm. The antenna has an operational band from 13.5 to 15.5 GHz, with a fractional bandwidth of 13.8 %. The antenna has one arm of the $\lambda/2$ dipole printed on one side of the substrate, with the other arm printed on the other. A printed reflector is added at the point of connection of the SMA to the substrate to improve the (Front to Back) F/B ratio.

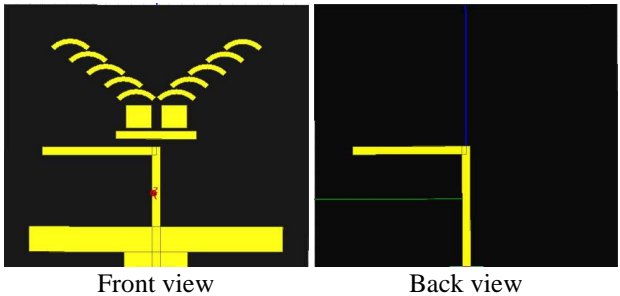

Front view      Back view
Figure 2. Antenna design view

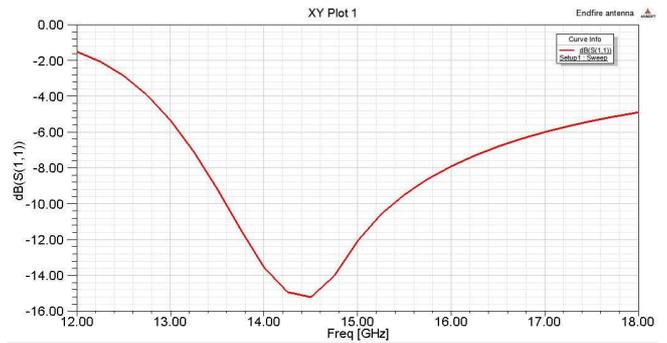

Figure 3. Return loss of the antenna

The radiation from the dipole being omni-directional is made directional by the reflector and by the director element, that couples the dipole electric field along the substrate plane. The surface waves are directed in the end fire direction aided by the in-plane reflector. This coupling is further enhanced by suitably spaced directors that channel the energy along two diagonal directions, aided by arc shaped transitions spaced at 0.01 $\lambda$ at 14 GHz. The return loss behavior is shown in Figure 3. It has a return loss < 10 dB in the band 13.5 GHz to 15.5 GHz.

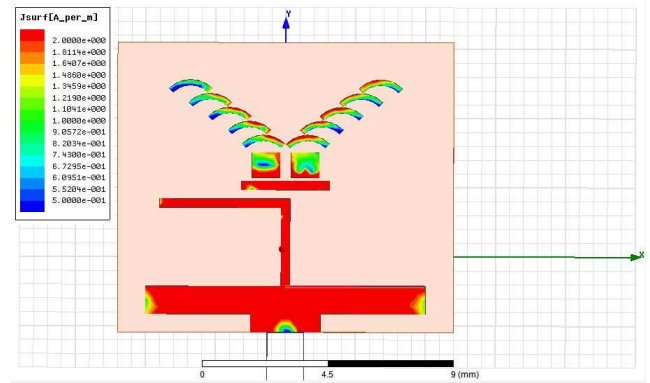

Figure 4. Surface current distribution at 14 GHz

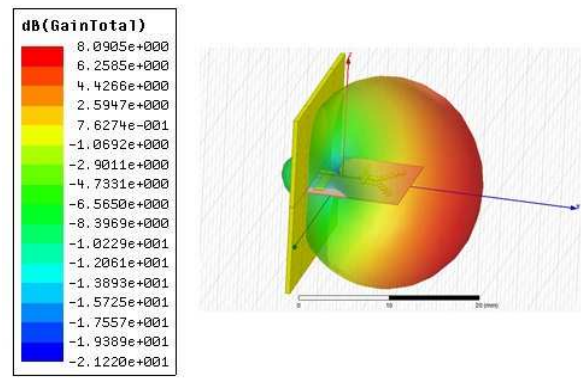

Figure 5. 3D antenna radiation pattern

Figure 4 shows the surface current distribution on the antenna at 14 GHz. The currents, as can be seen, are very strong on the dipole and resonate at 14 GHz for a 9 mm ($\lambda/2$) long arm length. The currents from these dipole arms couple strongly to the director and the two rectangular parasitic resonators. The arc shaped resonators are arranged in a flaring fashion, such that the distance between the last arc elements is $\lambda/4$. This arrangement enhances the gain in the end fire direction, due to a constructive interference of the surface waves. The gap between the arc shaped resonators is maintained at λ/10 for the best coupling and to increase the resonant bandwidth. The antenna resonates with a directional beam of 8.1 dBi and the pattern is shown in Figure 5.

## IV. RECONFIGURABLE HIGH GAIN MULTIBEAMS

Massive MIMO and millimeter wave antenna system setups would benefit from the use of both beamforming and beamswitching schemes. The following sections detail the beamsteering and beamswitching performance of the end fire element when used in a suitable array.

### A. Beamsteering

The endfire antenna element array was simulated for the beam steering angle limit before the onset of grating lobes. It was seen that the beam can be steered away from the boresight direction to a margin of 30 degrees, as shown in Figure 6. The pointing losses when the beam is steered are also depicted in the figure normalized w.r.t the boresight direction. The F/B ratio can be further improved by using an aluminium reflector behind the array.

For this simulation, a linear arrangement of 5 elements of the antenna was used with an inter-element spacing 0.6 λ - a little over the 0.5 λ rule to avoid grating lobes. The beam steering was tested by feeding the phases progressively in steps of 30 degrees to each of the individual antenna elements.

Beamsteering achieved by simulations in this fashion helps in gauging the performance of the antenna array to progressive phase shifts and the angular limit, to which the same can be applied while accounting for the trade-off between inter-element spacing and grating lobes.

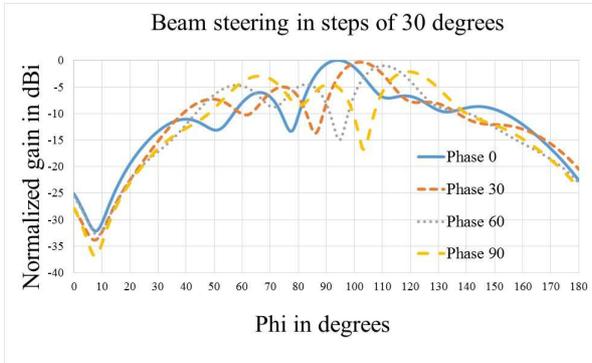

Figure 6. Beamsteering with progressive phase shift

### B. Beamswitching scheme

The single element end fire antenna is placed in a 3x3 planar array with a spacing of λ/2 at 14 GHz to avoid Grating lobes as shown in Figure 7. By using two different excitation phases to the elements of the antenna array, a reconfigurable high gain multibeam system is developed. The excitation to the individual antenna can be represented as matrix element of the 3x3 excitation matrix and this can be either a signal of amplitude k or -k, depending on the pattern to be excited. The extension of this scheme to a 4x4 array is depicted in Figure 8 along with the projected beams. The excitation matrices developed in the next section can by analogy be derived for a 4x4 subarray to obtain similar beamswitching performance. This will later be scaled to the 8x8 beamswitching scheme.

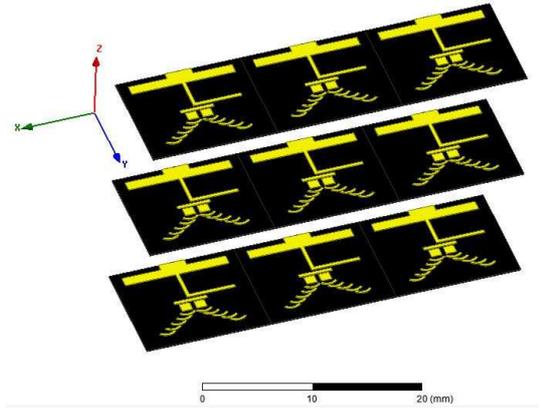

Figure 7. 3x3 endfire antenna array

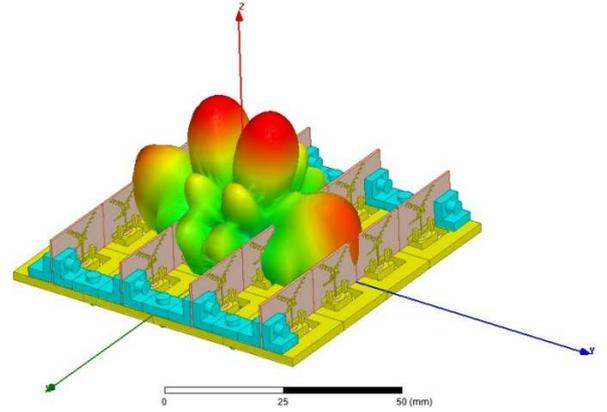

Figure 8. A 4x4 array projecting two simultaneous high gain beams using a scheme template deduced from Table 1

### C. Excitation matrices of the beamswitching scheme

The different matrices with the antenna excitations as individual elements of the matrix are shown below. It can be seen that these matrices are related to each other by simple matrix operations and therefore the switching between different excitations can be performed using simple FPGA techniques. For our purposes, we verify these operations using full wave simulations. A-F depict the different antenna beamstates.

$$A = \begin{bmatrix} k & k & k \\ k & k & k \\ k & k & k \end{bmatrix}$$

$$B = \begin{bmatrix} -k & 0 & -k \\ 0 & 0 & 0 \\ k & 0 & k \end{bmatrix} \quad C = \begin{bmatrix} -k & 0 & k \\ 0 & 0 & 0 \\ -k & 0 & k \end{bmatrix}$$

$$C = B^T$$

$$D = \begin{bmatrix} 0 & k & 0 \\ -k & 0 & -k \\ 0 & k & 0 \end{bmatrix} \quad E = \begin{bmatrix} 0 & k & 0 \\ k & 0 & -k \\ 0 & -k & 0 \end{bmatrix}$$

$$F = \begin{bmatrix} 0 & k & 0 \\ -k & 0 & k \\ 0 & -k & 0 \end{bmatrix}$$

Table 1 depicts the antenna state and gives a variation of the excitation vector over the 9 ports. It can be clearly seen that it is a two-vector variation.

| PORT NUMBERS | ANTENNA STATE | | | | | |
|---|---|---|---|---|---|---|
| | A | B | C | D | E | F |
| 1 | K | K | K | 0 | 0 | 0 |
| 2 | K | 0 | 0 | K | -K | -K |
| 3 | K | K | -K | 0 | 0 | 0 |
| 4 | K | 0 | 0 | -K | -K | K |
| 5 | K | -K | K | 0 | 0 | 0 |
| 6 | K | 0 | 0 | 0 | 0 | 0 |
| 7 | K | 0 | 0 | K | K | K |
| 8 | K | 0 | 0 | -K | K | -K |
| 9 | K | -K | -K | 0 | 0 | 0 |

Table 1. Pattern steering excitation. $k = 1\angle 0$, $-k = 1\angle 180$, ´0´ implies no port excitation

### D. Switched matrix antenna patterns

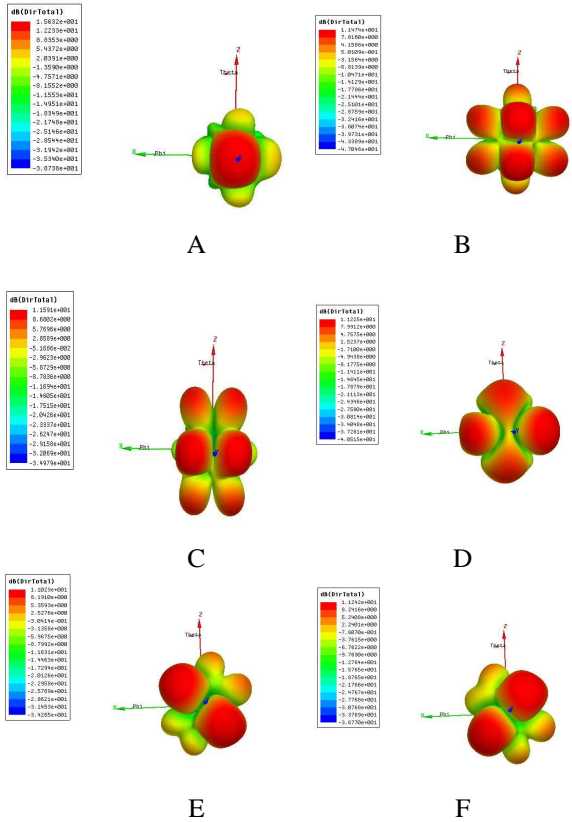

A  B
C  D
E  F

Figure 9. Antenna states with different radiation pattern at 14.5 GHz for the 3x3 array

Figure 9 depicts the antenna beamstates. Antenna state A depicts a boresight beam of high directivity emerging perpendicular to the array plane. Antenna state B and C represent two beams in the vertical and horizontal plane with simultaneous high gains. Antenna state D has 4 beams pointing along the four directions with simultaneous high gains. Finally, state E and F represent slanted beams in +45 and -45 directions. It is important to note that the antenna beam patterns were stable throughout the operating bandwidth of the antenna for the phase excitations.

## V. CONCLUSION

Massive MIMO antennas operating in the mmWave band require high gain antennas that can be beamsteered along different directions with simultaneous beams while casting nulls along other interference directions. The antenna element presented here is a potential high gain antenna element in this regard. The beam switching and beam steering potential of this antenna element in an antenna array was demonstrated using full wave simulations. The antenna can find applications also in mmWave beamforming for 5G cellular networks that require low form factor, high gain and reconfigurability in a small cell or femto cell scenario.

Further work would aim at the design of a feeding network that can efficiently handle these phases with low latency in switching. The actual prototype will be fabricated and the patterns measured in an anechoic chamber for the corresponding directionality of each antenna beamstate. The simultaneous multi-beam projection would enable the Massive MIMO scenario as was discussed in the paper. The high gain beamforming in millimeter wave frequencies using this concept will be taken up using the designed prototype. For Single RF MIMO different antenna beamstates operated over time (TDD), over the same link would be taken up- in tandem, with the research roadmap of achieving a synergy between the three Massive MIMO, millimeter wave and Single RF MIMO technologies.

## *Acknowledgment*

This work was supported by the 5Gwireless project that has received funding from the European Union's Horizon 2020 research and innovation programme under grant agreement No 641985.